\documentclass[11pt]{article}
\pdfoutput=1

\usepackage{bm,wrapfig,float,array,mathtools}
\usepackage{amsfonts}
\usepackage{amssymb}
\usepackage{cite}
\usepackage{amsmath}
\usepackage{color}
\usepackage{graphicx}
\usepackage{hyperref}
\usepackage{float}
\usepackage[T1]{fontenc}
\usepackage[utf8]{inputenc}
\usepackage{alphabeta}
\usepackage{fancyvrb}
\usepackage[dvipsnames]{xcolor}
\usepackage{bold-extra}
\usepackage{lmodern}
\usepackage{cleveref}
\usepackage[normalem]{ulem}
\usepackage{tikz-cd}

\graphicspath{ {figures/} }

\usepackage{tikz}
\usetikzlibrary{arrows}
\usetikzlibrary{arrows.meta}
\usetikzlibrary{shapes.geometric,calc,arrows, positioning,shapes.misc,decorations.markings}
\tikzset{
  big arrow/.style={
    decoration={markings,mark=at position 1 with {\arrow[scale=2,#1]{>}}},
    postaction={decorate},
    shorten >=0.4pt},
  big arrow/.default=black}
  
\pgfdeclarelayer{edgelayer}
\pgfdeclarelayer{nodelayer}
\pgfsetlayers{edgelayer,nodelayer,main} 
\tikzstyle{none}=[inner sep=0pt]

\tikzstyle{brane}=[draw]
\tikzset{D7/.style={circle, draw=black, inner sep=0pt, fill=white, minimum size=3mm}}
\tikzset{hasse/.style={circle, fill,inner sep=2pt}}
\tikzset{flavor/.style={regular polygon,regular polygon sides=4,inner sep=2.5pt, draw}}
\tikzset{gauge/.style={circle, draw,inner sep=2.5pt}}
\tikzset{gaugeb/.style={circle, draw,fill=black,inner sep=2.5pt}}
\tikzset{gaugecyan/.style={circle, draw,fill=cyan,inner sep=2.5pt}}
\tikzset{gaugegreen/.style={circle, draw,fill=green,inner sep=2.5pt}}
\tikzset{gaugeblue/.style={circle, draw,fill=blue,inner sep=2.5pt}}
\tikzset{gaugeorange/.style={circle, draw,fill=orange,inner sep=2.5pt}}
\tikzset{bd/.style={circle, draw=black, inner sep=0pt, fill=black, minimum size=2mm}}
\tikzset{wd/.style={circle, draw=black, inner sep=0pt, fill=white, minimum size=2mm}}
\tikzset{Dynkin/.style={circle, draw=black, inner sep=0pt, fill=white, minimum size=2mm}}
\tikzstyle{ligne}=[draw, thick] 
\tikzset{doublearrow/.style={ draw=black!75, color=black!75, thick, double distance=3pt, }}

\tikzstyle{NodeCross}=[draw, shape=circle, cross out, inner sep=0pt, minimum size=6pt,line width=0.25mm]
\tikzstyle{Circle}=[draw, shape=circle, black, inner sep=0pt, minimum size=6pt]
\tikzstyle{rtriangle}=[fill=black, regular polygon, regular polygon sides=3, rotate=90, inner sep=0pt, minimum size=8pt]
\tikzstyle{ltriangle}=[fill=black, regular polygon, regular polygon sides=3, rotate=270, inner sep=0pt, minimum size=8pt]
\tikzstyle{rtriangleblue}=[fill={rgb,255: red,17; green,160; blue,255}, regular polygon, regular polygon sides=3, rotate=90, inner sep=0pt, minimum size=8pt]
\tikzstyle{ltriangleblue}=[fill={rgb,255: red,17; green,160; blue,255}, regular polygon, regular polygon sides=3, rotate=270, inner sep=0pt, minimum size=8pt]
\tikzstyle{rtrianglegreen}=[fill={rgb,255: red,69; green,255; blue,28}, regular polygon, regular polygon sides=3, rotate=90, inner sep=0pt, minimum size=8pt]
\tikzstyle{ltrianglegreen}=[fill={rgb,255: red,69; green,255; blue,28}, regular polygon, regular polygon sides=3, rotate=270, inner sep=0pt, minimum size=8pt]
\tikzstyle{Uprtriangle}=[fill=black, regular polygon, regular polygon sides=3, rotate=0, inner sep=0pt, minimum size=8pt]
\tikzstyle{Downltriangle}=[fill=black, regular polygon, regular polygon sides=3, rotate=180, inner sep=0pt, minimum size=8pt]
\tikzstyle{rtriangleAmber}=[fill={rgb,255: red, 191; green, 144; blue, 63}, regular polygon, regular polygon sides=3, rotate=90, inner sep=0pt, minimum size=8pt]
\tikzstyle{UprtriangleViolett}=[fill={rgb,255: red,255; green,0; blue,0}, regular polygon, regular polygon sides=3, rotate=0, inner sep=0pt, minimum size=8pt]
\tikzstyle{ltrianglered}=[fill={rgb,255: red,191; green,0; blue,0}, regular polygon, regular polygon sides=3, rotate=270, inner sep=0pt, minimum size=8pt]

\tikzstyle{Downltriangle}=[fill=black, regular polygon, regular polygon sides=3, rotate=180, inner sep=0pt, minimum size=8pt]
\tikzstyle{UpRighttriangle}=[fill=black, regular polygon, regular polygon sides=3, rotate=45, inner sep=0pt, minimum size=8pt]
\tikzstyle{UpLefttriangle}=[fill=black, regular polygon, regular polygon sides=3, rotate=315, inner sep=0pt, minimum size=8pt]
\tikzstyle{DownRighttriangle}=[fill=black, regular polygon, regular polygon sides=3, rotate=135, inner sep=0pt, minimum size=8pt]
\tikzstyle{DownLighttriangle}=[fill=black, regular polygon, regular polygon sides=3, rotate=225, inner sep=0pt, minimum size=8pt]

\tikzstyle{Star}=[draw, shape=star, fill=black, star points=8, inner sep=0pt, minimum size=8pt]

\tikzstyle{DashedLine}=[-, densely dashed, line width=0.25mm]
\tikzstyle{DashedLineBrown}=[-, densely dashed, line width=0.25mm, draw={rgb,255: red,155; green,103; blue,51}]
\tikzstyle{DashedLineFall}=[-, densely dashed, line width=0.25mm, draw={rgb,255: red,195; green,0; blue,0}]
\tikzstyle{DashedLineViolett}=[-, densely dashed, line width=0.25mm, draw={rgb,255: red,139; green,41; blue,148}]
\tikzstyle{DottedLine}=[-, dotted, line width=0.25mm]
\tikzstyle{BlueLine}=[-, fill=none, draw={rgb,255: red,17; green,160; blue,255}, line width=0.25mm]
\tikzstyle{GreenLine}=[-, fill=none, draw={rgb,255: red,69; green,255; blue,28}, line width=0.25mm]
\tikzstyle{RedLine}=[-, draw={rgb,255: red,191; green,0; blue,0}, fill=none, line width=0.25mm]
\tikzstyle{LBRedLine}=[-, draw={rgb,255: red,255; green,0; blue,0}, fill=none, line width=0.25mm]
\tikzstyle{DashedLineRed}=[-, densely dashed, fill=none, draw={rgb,255: red,191; green,0; blue,0}, line width=0.25mm]
\tikzstyle{ThickLine}=[-, line width=0.25mm]
\tikzstyle{ViolettLine}=[-, draw={rgb,255: red,132; green,60; blue,191}, fill=none, line width=0.25mm]
\tikzstyle{ViolettDashedLine}=[-, densely dashed, draw={rgb,255: red,132; green,60; blue,191}, fill=none, line width=0.25mm]
\tikzstyle{AmberLine}=[-, draw={rgb,255: red,191; green,144; blue,63}, fill=none, line width=0.25mm]
\tikzstyle{DashedRedThick}=[-, densely dashed, fill=none, draw={rgb,255: red,191; green,0; blue,0}, line width=0.40mm]
\tikzstyle{DashedBlueThick}=[-, densely dashed, fill=none, black, line width=0.40mm]
\tikzstyle{DottedLineRed}=[-, dotted, line width=0.25mm, draw={rgb,255: red,191; green,0; blue,0}]
\tikzstyle{DottedLineBlue}=[-, dotted, line width=0.25mm, draw={rgb,255: red,17; green,160; blue,255}]
\tikzstyle{ArrowLineRight}=[-, -{Stealth[scale=1.75]}, line width=0.1mm, scale=5]
\tikzstyle{ArrowLineLeft}=[-, {Stealth[scale=1.75]}-, line width=0.1mm, scale=5]
\tikzstyle{ArrowLineRightBlue}=[-, -{Stealth[scale=1.75]}, line width=0.1mm, draw={rgb,255: red,17; green,160; blue,255}]
\tikzstyle{ArrowLineLeftBlue}=[-, {Stealth[scale=1.75]}-, line width=0.1mm, draw={rgb,255: red,17; green,160; blue,255}]
\tikzstyle{ArrowLineRightRed}=[-, -{Stealth[scale=1.75]}, line width=0.1mm, draw={rgb,255: red,191; green,0; blue,0}]
\tikzstyle{ArrowLineLeftRed}=[-, {Stealth[scale=1.75]}-, line width=0.1mm, draw={rgb,255: red,191; green,0; blue,0}]

\newcommand{\bea}{\begin{eqnarray}}
\newcommand{\eea}{\end{eqnarray}}
\newcommand{\be}{\begin{equation}}
\newcommand{\ee}{\end{equation}}
\newcommand{\ba}{\begin{aligned}}
\newcommand{\ea}{\end{aligned}}
\newcommand{\bit}{\begin{itemize}}
\newcommand{\eit}{\end{itemize}}
\newcommand{\ben}{\begin{enumerate}}
\newcommand{\een}{\end{enumerate}}

\begin{document}

\baselineskip=18pt  
\numberwithin{equation}{section}  
\allowdisplaybreaks  

\thispagestyle{empty}

\vspace*{0.8cm} 
\begin{center}
{{\Huge  \bf{Higher Form Symmetries TFT in 6d}}}

 \vspace*{1.5cm}
Fabio Apruzzi $^\flat$

 \vspace*{.5cm} 
{\it $^\flat$Albert Einstein Center for Fundamental Physics, Institute for Theoretical Physics,\\
University of Bern, Sidlerstrasse 5, CH-3012 Bern, Switzerland}\\

\vspace*{0.8cm}
\end{center}
\vspace*{.5cm}

\noindent
Symmetries and anomalies of a $d$-dimensional quantum field theory are often encoded in a $(d+1)$-dimensional topological action, called symmetry topological field theory (TFT). 
We derive the symmetry TFT for the 2-form and 1-form symmetries of 6d $(1,0)$ field theories, focusing on theories with a single tensor multiplet (rank 1). We implement this by coupling the low-energy tensor branch action to the background fields for the higher-form symmetries and by looking at the symmetry transformation rules on dynamical and background fields. These transformation rules also imply a mixing of the higher-form symmetries in a 3-group structure. For some specific and related higher rank cases, we also derive the symmetry TFT from the holographic dual IIA supergravity solutions. The symmetry TFT action contains a coupling between the 2-form symmetry and the 1-form symmetry backgrounds, which leads to a mixed anomaly between the 1-form symmetries of the 5d KK-theory obtained by circle compactification. We confirm this by a pure 5d analysis provided by the 5d effective low-energy Coulomb branch Lagrangian coupled to background fields. We also derive the symmetry TFT for 5d $SU(p)$ supersymmetric gauge theories with Chern-Simons level $q$ and for 5d theories without non-abelian gauge theory description at low-energy. Finally, we discuss the fate of the 2-form and 1-form symmetry of rank 1 6d field theories when coupled to gravity.

\newpage

\tableofcontents

\newpage

\section{Introduction}
Symmetries are key in quantum field theory and they constrain the spectrum of states and local operators. The notion of symmetry has been extended to include generalized higher form symmetries \cite{Gaiotto:2014kfa}, which constrain the spectrum of extended objects like for instance Wilson and 't Hooft lines of a gauge theory. A generalized $p$-form symmetry, $\Gamma^{(p)}$, is generated by a (charge) topological operator $U_g$ supported on a $(d-p-1)$-dimensional subspace, $M^p$ of $d$-dimensional space-time, and it acts on a $p$-dimensional object (when $p>0$ these are sometimes called defects), which is defined as the charge object. The standard case is when $p=0$, where the charged object are point particle operators $\mathcal{O}$, whereas the charge operator $U_g$ is supported on a $(d-1)$-dimensional manifold  surrounding $\mathcal{O}$. In general, we have that the linking of the topological operator with the charged one gives the symmetry group element, $g(\mathcal{O})$, as follows,
\begin{equation}
U_g(M^{d-p-1}) \mathcal{O}(M^{p}) = g(\mathcal{O})  \mathcal{O}(M^{p})U_g(M^{d-p-1}).
\end{equation}
Higher form symmetries can be both continuous and discrete but they are abelian by construction. 

These generalized symmetries are present in many physical system of interest to condensed matter and particle physics, from $d=2$ space-time dimensions \cite{Pantev:2005rh} to $d=4$. In particular, 1-form symmetries can be present in gauge theories. Gauge theories with adjoint or no matter have discrete 1-form symmetries corresponding to the center of the gauge group $Z(G)$, and they act on line operators of the gauge theory, which can be Wilson or 't Hooft lines. In addition, 1-form symmetries depend on and are closely related to the global structure of the gauge group as seen in \cite{Aharony:2013hda, Gaiotto:2014kfa}. The presence of matter fields can break the 1-form symmetries by screening, that is a phenomenon consisting in the matter particles ending on the line operator, making the topological operator that define the symmetry, $U_g$, trivial. 1-form symmetries are present in gauge theories in any dimensions from $d=2,3,4$ \cite{Benini:2017dus, Hsin:2018vcg, Cherman:2019hbq, Gaiotto:2017yup, Anber:2015wha} to $d>4$ \cite{Morrison:2020ool, Albertini:2020mdx, Apruzzi:2020zot, Bhardwaj:2020phs}. In Higher dimensions $(d>4)$, Lagrangian field theories can only be effective low-energy descriptions of interacting strongly coupled field theories, whose existence has been predicted via string theory geometric constructions, such as superconformal field theories (SCFTs).
In particular, 6d $(1,0)$ and $(2,0)$ theories have also 2-form symmetries due to the presence of dynamical anti-symmetric tensor fields \cite{Gaiotto:2014kfa, DelZotto:2015isa}.
The knowledge of standard and generalized symmetries provide a useful tool to constrain local and extended objects of higher dimensional field theories as well. Moreover, string theory together with geometric engineering provide frameworks where all possible symmetries can be computed from the boundary geometry \cite{Apruzzi:2018nre, Apruzzi:2019opn, Apruzzi:2021vcu, Bhardwaj:2020phs, Bhardwaj:2020ruf, Morrison:2020ool, Albertini:2020mdx, GarciaEtxebarria:2019caf}.

Generalized $p$-form symmetries can have 't Hooft anomalies, which are robust quantities expressed in terms of the background fields of these symmetries. One or more symmetries might be involved, in this case the anomaly is called mixed. In particular, they are seen as the obstruction for gauging all the involved symmetries. If an effective Lagrangian description exists, these anomalies can be computed by coupling the theory to the background field for the $p$-form symmetry, that is a $(p+1)$-form gauge field. 't Hooft anomalies are quantized quantities, which do not depend on any scale and therefore they are robust under RG-flow. For this reason, they have been importantly used to constrain the strongly coupled dynamics of interacting quantum field theory via what is indeed called anomaly matching. For instance, in 4d adjoint QCD theories there are anomalies between the chiral symmetry (which acts on the fermions and the $\theta$-parameter) and the 1-form constrains the infra-red (IR) physics \cite{Gaiotto:2014kfa, Gaiotto:2017yup, Cordova:2018acb, Cordova:2019jnf, Cordova:2019uob, Cordova:2019bsd, Cox:2021vsa}, and predict certain non-trivial features of the strongly coupled regime, which are difficult to access otherwise. Anomalies of generalized symmetries are present also in higher $d>4$ dimensions. In 5d supersymmetric $SU(N)$ gauge theories with no massless matter and with possibly a classical Chern-Simons level \cite{Intriligator:1997pq}, there is a mixed anomaly between the instanton symmetry, given by the following topological current $J_I=\ast_5 \frac{1}{4}\text{Tr}(f\wedge f)$ and the 1-form symmetry $\mathbb{Z}_N$ \cite{BenettiGenolini:2020doj}. When the classical Chern-Simons level is present, there is also a cubic 't Hooft anomaly for the 1-form symmetry \cite{Gukov:2020btk}.

Symmetries, anomalies and the choice of global structure of a $d$-dimensional theory can be encoded in a $(d+1)$ topological field theory action \cite{Freed:2012bs, Freed:2014iua, Witten:1998wy, Aharony:1998qu, Gross:1998gk, Bergman:2020ifi, Bah:2020uev, Apruzzi:2021phx}. This $(d+1)$-dimensional theory is called symmetry topological field theory (Symmetry TFT or SymTFT). String theory and geometric engineering provide a systematic framework to compute the symmetry TFT of the engineered models via tools described and developed in \cite{Apruzzi:2021nmk}. 5d SCFTs are engineered by 3-dimensional Calabi-Yau cones in M-theory or IIB five-brane webs. As described in \cite{Apruzzi:2021nmk}, the symmetry TFT for these theories has been derived by reducing the topological couplings of 11d supergravity on the boundary of the Calabi-Yau cones or from the IIB webs.

In this paper we derive the symmetry TFT involving 1-form symmetries and 2-form symmetries of 6d $(1,0)$ SCFTs from their tensor branch description, by using the bosonic Lagrangian of the theory when the scalar components of the tensor multiplets acquire non-vanishing vacuum expectation value (vev). This is done by generalizing to higher dimensions the 4d Maxwell theory case analyzed in \cite{Gaiotto:2014kfa, Cordova:2018cvg}. The SymTFT is derived by coupling the Lagrangian to the higher-form symmetry backgrounds, and by analyzing the transformation rules of these symmetries on the various dynamical and background fields. A first consequence of the 1-form symmetry transformation is that the background field for the 2-form symmetry should transform accordingly, i.e. by mixing with 1-form symmetry transformations in order to avoid dangerous ambiguities that depend on dynamical fields. This signals the presence of a 3-group structure. Moreover, some ambiguities cannot be reabsorbed by counterterms, but rather by a 7d topological action. This is the symmetry TFT for the 6d theories, and in this case is not invertible, which means that the theory living at the boundary does not have a partition function but rather a partition vector (see \cite{Freed:2012bs, Freed:2014iua} for the definition of invertible TFT). In general, there can be boundary conditions leading to a theory at the boundary that is absolute, i.e. with a well defined partition function. This symmetry TFT contains also a coupling between the 2-form and the 1-form symmetry backgrounds. We then discuss explicit examples, focusing on (tensor branch) rank 1 theories. We also study absolute theories coming from these rank 1 SCFTs, and combinations thereof. The nontrivial coupling between the 2-form and 1-form symmetry of the symmetry TFT does not lead to a mixed anomaly for any of the absolute theories studied. To support this we provide a derivation of the symmetry TFT for some specific examples constructed holographically via the AdS$_7 \times M_3$ solutions of IIA supergravity in \cite{Apruzzi:2013yva, Apruzzi:2017nck} with orientifold O6$^+$ planes sources with D6 branes on top. This is done by reducing on $M_3$ the topological coupling of IIA supergravity, including Chern-Simons couplings related to the brane sources, see \cite{Witten:1998wy, Aharony:1998qu, Gross:1998gk, Belov:2004ht, Maldacena:2001ss, Bergman:2020ifi, Apruzzi:2021phx} for similar examples.
Subsequently, we study the compactification on $S^1$ of this symmetry TFT on a circle. This consists of the symmetry TFT for the 5d $\mathcal{N}=1$ kk-theory (which uplift to the 6d $(1,0)$ SCFT in the UV). In this case there are choices of boundary conditions which lead to absolute theories with an invertble symmetry TFT, leading to anomalies of the 5d kk-theory, or alternatively giving rise to a theory with a 3-group. We also derive the symmetry TFT from the 5d low-energy Coulomb branch action coupled to backgrounds, generalizing the procedure and results of \cite{Cvetic:2021sxm}. As a bonus we also recover the symmetry TFT for 5d $SU(p)_q$ gauge theories with $q$ Chern-Simons level and for the ${B_N,B_N^{(1)},B_N^{(2)}}$ discussed in \cite{Eckhard:2020jyr, Morrison:2020ool, Apruzzi:2021nmk}, which do not have a non-abelian gauge theory effective description at low-energy. Finally we comment on the fate of the 2-form symmetries of 6d rank 1 theories when we couple them do 6d dynamical gravity. Compatibly with \cite{Banks:2010zn} all global symmetries must be either gauged or broken. We provide evidence that when the 2-form symmetry is broken the surface defect charged under them are screened by the supergravity strings. This is detected by checking whether the Dirac quantization condition for the supergravity strings is violated or not \cite{Apruzzi:2020zot}. In other examples the 2-form symmetry survives \cite{Braun:2021sex} as well as the diagonal combination of 1-forms symmetries. Therefore the full 3-group including the 1-form symmetry part should be gauged. 

The paper is organized as follows. In section \ref{sec:symtft6d} we derive the symmetry TFT for 6d $(1,0)$ relative theories from their tensor branch, and we discuss what survives for absolute ones. In section \ref{sec:holoTFT} we derive the symmetry TFT holographically. In section \ref{sec:5dperp} we discuss the 5d perspective on these SymTFT deriving the circle compactification and directly from the 5d Coulomb branch effective action. Finally in section \ref{sec:sugra} we discuss implication for 2 and 1-form symmetries in 6d when coupled to dynamical gravity.

\section{2-form and 1-form symmetries TFT from the tensor branch} \label{sec:symtft6d}
We start by deriving the Symmetry TFT for 2-form and 1-form symmetries of a generic $\mathcal{N}=(1,0)$ SCFTs in the tensor branch.  A tensor branch (pseudo)-Lagrangian description comes from taking the vev of the scalar component of the tensor multiplet such that $\langle\phi^i\rangle \neq 0$.  The theory consists of $N_T$ tensor multiplets,  coupled to a quiver gauge theory.  For each tensor multiplet there is an associated gauge group, which can also  be trivial. The gauge groups are connected by bifundamental hypermultiplets, forming a quiverlike structure.  Moreover there can be matter (hypermultiplets) rotating under certain flavor groups. The generic bosonic pseudo-action reads\footnote{Where we used the notation for which the instanton density of $SU(N)$, $\frac{1}{4}{\rm Tr}(f\wedge f)$, has integer periods. Therefore we rescaled $f$ by $2\pi$ and {\rm Tr} is the normalized trace ${\rm Tr}(f\wedge f)=  \frac{{\rm tr}_{\rm fund}(f\wedge f)}{{\rm Ind}({\rm fund}(SU(N)))}$, with ${\rm Ind}({\rm fund}(SU(N)))= \frac{1}{2}$.}
\begin{equation} \label{eq:TBact}
S \supset 2\pi \int \Omega_{ij} \big( -\tfrac{1}{2}d\phi^i \wedge \ast d\phi^j - \tfrac{1}{4}h^i \wedge \ast h^j \big) + \Omega_{ij} \big( \phi^i \wedge \tfrac{1}{4}\text{Tr}(f^j \wedge \ast f^j)+ b^i \wedge \tfrac{1}{4}\text{Tr}(f^j \wedge f^j)\big)
\end{equation}
where $\phi^i$ and $b^i$ are the scalars and antisymmetric 2-form fields of the dynamical tensor multiplets,  whereas $f^i$ are the field strength of the gauge vectors.  The coupling is dictated by the Dirac pairing,  $\Omega_{ij}$,  in the BPS integral string lattice of charges under $b^i$.  Finally,  we recall that \eqref{eq:TBact} is a pseudo-action where we need to impose the (anti-)self duality constraint on $ db^i = \pm \ast_6 db^i$.
The Bianchi identity reads $dh^i =  \tfrac{1}{4}\text{Tr}(f^i \wedge f^i)$,  such that 
\begin{equation}
h^i = db^i - \kappa CS_3(a^i)
\end{equation}
for some integral Chern-Simons coefficient $\kappa$.
Therefore there are $N_T$ 2-form conserved currents given by $J^i_{(2)}=\ast_6 db_i$. 

We now implement a similar strategy to \cite{Cordova:2018cvg}. We couple the theory to the background fields for the 2-form symmetries associated to each $U(1)$ two-form gauge fields, $b^i$. We denote them by $C_3^i$. The bosonic action coupled to the 2-form symmetries background reads,
\begin{equation}
S[C_3^i]= 2\pi \int \Omega_{ij}  \left( - \tfrac{1}{4}(h^i - C_3^i) \wedge \ast(h^j - C_3^j) +  b^i \wedge \tfrac{1}{4}\text{Tr}(f^j \wedge f^j) + db^i \wedge C_3^j -  \kappa CS_3(a^i) \wedge  C_3^j  \right)
\end{equation}
where we ignored the part involving the scalars and the kinetic term is invariant under the 2-form symmetry transformations,
\begin{equation} \label{eq:2fstransf}
b^i \rightarrow b^i + \Lambda_{(2)}^i,  \qquad C_3^i \rightarrow C_3^i +d \Lambda_{(2)}^i
\end{equation}
the third term is  the coupling of $C_3^i$ to the 2-form symmetry currents $J^i_{(2)}$,  and the last one is a counterterm.  This is the very low-energy effective action together with its symmetries and the backgrounds thereof.  Accounting for the BPS strings charged under the dynamical $b^i$, the 2-form symmetry is generically broken to the following product
\begin{equation}
\Gamma^{(2)} =\prod_i \mathbb{Z}_{n_i}
\end{equation}
which is dictated by the Smith normal form of the pairing in the string charge lattice $\Omega_{ij}$, that is generically $\Omega_{\rm SNF}= {\rm diag}\{n_1, \ldots, n_i, \ldots\}$ \cite{Bhardwaj:2020phs}.

6d theories, more specifically 6d SCFTs\footnote{6d SCFTs do not have continuous 1-form symmetries, \cite{Cordova:2020tij}.} on the tensor branch, can also have 1-form symmetries, which sit in the center of the gauge groups (let us suppose we have a single one) \cite{Apruzzi:2020zot, Bhardwaj:2020phs, Apruzzi:2021mlh}. Moreover, because of the absence of massless gauge $U(1)$ \cite{Hanany:1996ie} the 1-form symmetry that sit in the center of the gauge group can be at most discrete. Coupling the theory to its $B_2$ background field we also have an additional term that is,
\begin{equation}
S[C_3^i, B_2^i] = S[C_3^i] + 2\pi \int \Omega_{ij}  b^i \, \alpha_G^j \mathfrak{P}(B_2)
\end{equation}
where $\mathfrak{P}(B_2)$ is the Pontryagin square due to the fractionalization of the instanton number when activating $B_2$ that sits in the center of the gauge groups \cite{Cordova:2019uob}.  Moreover,  $\alpha_G^j $ are generically fractional coefficient that depend on the gauge groups as well as which subgroup of the associated center symmetry has been activated.  

We can look now what happens when we apply the 2-form symmetry transformation \eqref{eq:2fstransf}.  The action shift as follows,
\begin{equation}
S[C_3^i, B_2^i]  \rightarrow S[C_3^i, B_2^i] +  2\pi \int  \Omega_{ij} \left(\frac{1}{2} d\Lambda_{(2)}^i \, C_3^j + \Lambda_{(2)}^i \, \alpha_G^j \mathfrak{P}(B_2) \right)
\end{equation}
This shift cannot be reabsorbed by adding additional counterterms, but instead can be reproduced by the following 7d action evaluated on the 6d boundary
\begin{equation} \label{eq:symTFT}
S_{\rm SymTFT} = 2\pi  \Omega_{ij} \int  \left( \frac{1}{2} C_3^i \, dC_3^j + C_3^i \, \alpha_G^j \mathfrak{P}(B_2) \right)
\end{equation}
This action is the key to read off the true 1-form symmetry of the theory, i.e.  by looking at boundary conditions of this action. Let us assume the theory has a $\Gamma^{(1)}=\mathbb{Z}_{p}$ symmetry, whose background is $B_2$. In order to avoid dangerous ambiguities, which depend on dynamical fields, due to symmetry transformations, we need to impose  
\begin{equation} \label{eq:3group}
B_2\rightarrow B_2^j + d\Lambda_{(1)}^j, \qquad C_3^j \rightarrow C_3^j +d\Lambda_{(2)}^j - p \alpha_G \Lambda_{(1)}^j B_2- p^2 \alpha_G \Lambda_{(1)}^j d\Lambda_{(1)}^j 
\end{equation}
This signals that the $\Gamma^{(2)}$ and $\Gamma^{(1)}$ combine in a 3-group structure, $Z_3$, via a generalized short exact sequence implementing the extension $1\rightarrow \Gamma^{(2)} \rightarrow Z_3 \rightarrow \Gamma^{(1)} \rightarrow 1$. Something similar happens in 4-dimension with the theta angle promoted to a dynamical axion field, \cite{Seiberg:2018ntt, Brennan:2020ehu}, since the dynamical $b^i$ are higher-form version of an 4d axion field in 6d. This structure will proliferate also in the explicit examples of the next subsection.

\subsection{Examples: rank 1 6d SCFTs}
For the sake of concreteness, let us analyze the 6d SCFTs with a single tensor multiplet. \cite{Bertolini:2015bwa}.  The cases with nontrivial 1-form symmetry \cite{Apruzzi:2020zot,Bhardwaj:2020phs,Apruzzi:2021mlh},  $\Gamma^{(1)}$, are 
\begin{align} \label{eq:NHCs}
\begin{tabular}{|c|c|c|c|c|}
\hline $\Omega_{ii}$ & $G$  & $F$ & $\Gamma^{(1)}$ & $\alpha_G$\\ \hline
$3$ & $SU(3)$ & $\varnothing$ & $\mathbb{Z}_3$ & $\frac{1}{3}$ \\ \hline
$4$ & $Spin(8)$ & $\varnothing$ & $\mathbb{Z}_2 \times \mathbb{Z}_2 $  & $\left(\frac{1}{2},\frac{1}{2}\right)$ \\ \hline
$4$ & $Spin(2k+4)$, $k>2$ & $Sp(2k-4)$ & $\mathbb{Z}_2$ & $\frac{1}{2}$ \\ \hline
$6$ & $E_6$ & $\varnothing$ &  $\mathbb{Z}_3$ & $\frac{2}{3}$\\ \hline
$8$ & $E_7$ & $\varnothing$ & $\mathbb{Z}_2$ & $\frac{1}{2}$ \\ \hline
\end{tabular}
\end{align}
where for $Spin(8)$ we have two $\alpha_G$ referring to the coefficients of $\mathfrak{P}(B_L+B_R)$ and $B_L\cup B_R$ respectively,  and $B_L, B_R$ are the background 1-form symmetry fields for the two $\mathbb{Z}_2$ factors. $F$ denotes the flavor symmetry rotating matter hypermultiplets when they are present. We also restrict to the global structure choice for the gauge group that leads to the simply connected case and electric 1-form symmetry. The symmetry TFTs are determined by plugging in the data \eqref{eq:NHCs} into \eqref{eq:symTFT}, for instance we have
\begin{equation}
S_{\rm SymTFT} = 2\pi  \int  \left( \frac{1}{2}\,  n \, C_3 \, dC_3+ 3\, n\, C_3 \,  \alpha_G \mathfrak{P}(B_2) \right)
\end{equation}
with $n=3,4,6,8$.
Specifying for concreteness to $n=3$ the SymTFT reads,
\begin{equation}
S_{\rm SymTFT} = 2\pi  \int  \left( \frac{1}{2}\,  3 \, C_3 \, dC_3+ 3\, C_3 \, \frac{1}{3} \mathfrak{P}(B_2) \right)
\end{equation}
where the 3-group symmetry transformation reads, 
\begin{equation}
    C_3\rightarrow C_3 + d\Lambda_{(2)} - \Lambda_{(1)} B_2- 3 \Lambda_{(1)} d\Lambda_{(1)}.
\end{equation}
For all these cases, the symmetry TFT does not describe an absolute quantum field theory living at the boundary,  i.e.  such that it has a well defined partition function,  but rather a relative field theory,  which instead has a partition vector,  see \cite{DelZotto:2015isa, Bhardwaj:2020phs, Morrison:2020ool, Albertini:2020mdx, Gukov:2020btk} for explicit example of relative vs absolute field theories. The important aspect is that a relative theory is described by a non-invertible TQFT in the bulk, whereas the anomalies of an absolute theory are described by invertible TQFTs. 

We now investigate whether the second term in \eqref{eq:symTFT}, which can potentially lead to an invertible TQFT for absolute theories, describing therefore a mixed anomaly between the 1-form symmetry,  $\Gamma^{(1)}$, and the 2-form symmetry,  $\Gamma^{(2)}$.  Many 6d $(1,0)$ SCFTs coming from string theory constructions are actually relative theories.  For instance, the rank 1 theories listed in \eqref{eq:NHCs} are all relative apart from the case with $\Omega_{ii}=n=p^2=4$,  with $p=2$.  Similarly to the $(2,0)$ case as discussed in \cite{Gukov:2020btk},  we can add a topological boundary term,
\begin{equation}
S_{\rm SymTFT} \rightarrow S_{\rm SymTFT} +  \frac{2 \pi}{2} \int_{\partial} 2 (C_3-B_3) Y
\end{equation}
where $B_3$ is a $\mathbb{Z}_2$ valued field living on the boundary,  and $Y$ is a $U(1)$ valued 3-form field.  The variation with respect to $Y$ results in 
\begin{equation}
C_3|_{\partial}= B_3 
\end{equation}
that is a boundary condition consistent with gauge invariance, leading to $L=\mathbb Z_2$ as maximal isotropic subgroup.  For these cases,  the second term in \eqref{eq:NHCs} has always integer coefficient,  due to $\oint C_3 \in \mathbb{Z}/2$,  and it does not lead to an anomaly term for an absolute 6d SCFT.  As in  \cite{Gukov:2020btk},  the only anomaly theory that is left is
\begin{equation}
S_{\rm an} = \frac{2 \pi}{2}  \int 4 B_3 d B_3.
\end{equation}

We can also take the product of two theories and understand what boundary condition are allowed and what are the maximal isotropic subgroup that survive as 2-form symmetries for absolute theories.  Let us focus on a product of two copies of the same rank 1 theory in \eqref{eq:NHCs}.  The maximal isotropic subgroup for the product of two $\Omega_{11}=4$ theories is isomorphic to $L=\mathbb{Z}_2$  \cite{Gukov:2020btk},  and therefore since $\oint C_3 \in \mathbb{Z}/2$ the second term in \eqref{eq:NHCs} has integral coefficient.  This works similarly in the case of $\Omega_{11}=8$.  So we conclude that the $C_3^i \wedge \alpha_G^j \mathfrak{P}(B_2)$ coupling does not lead to an anomaly for any absolute 6d $(1,0)$ theories studied in this section, i.e. coming from rank 1 theories or product thereof. It would be interesting to study higher rank cases,  which by a change of basis in the string charge lattice (or by a 7d generalization of the level rank duality \cite{Hsin:2016blu}) \cite{Gukov:2020btk},  can be understood in terms of a single $C_3$ field theory and its 7d topological action, that should correspond to the nontrivial ($\neq 1$) diagonal entries in $\Omega_{\rm SNF}$ \cite{Bhardwaj:2020phs}. We will see though how this turns out to be an anomaly under when we compactify the theory on $S^1$.

\section{SymTFT from holography} \label{sec:holoTFT}
Since 1-form symmetries are rare in 6d, let us discuss a specific example which has one, and it is a straightforward generalization of the rank 1 $\Omega_{11}$ with $G=Spin(2k+4)$. 
The IIA brane system which describe this theory is
\begin{equation}\label{eq:confbrane}
 \begin{tikzpicture}[baseline]
\node[circle,draw,fill=black,minimum size = .2cm,label=above: \footnotesize{$4N$\, NS5}] (n1) at (4,1) {};

\draw[solid,gray,very thick] (1.5,1)--(4-0.2,1) node[black,midway, xshift =0cm, yshift=0.3cm] {\footnotesize $\mathfrak{sp}(k-2)$} ;  
\draw[solid,gray,very thick] (4+0.2,1)--(6.5,1) node[black,midway, xshift =0cm, yshift=0.3cm] {\footnotesize $\mathfrak{sp}(k-2)$};

\draw [dashed,blue,very thick] (1.5,0.915)--(4-0.2,0.915) node[black,midway, xshift =0cm, yshift=-.3cm]{\footnotesize O6$^+$/D6}; 
\draw [dashed,blue,very thick] (4+0.2,0.915)--(6.5,0.915) node[black,midway, xshift =0cm, yshift=-.3cm]{\footnotesize O6$^+$/D6};
\end{tikzpicture} 
\end{equation}
where the 6-directions $x_0$ to $x_5$ of the 10d space of IIA supergravity are in common. The horizontal direction is $x_6$ and the vertical direction on the plane where \eqref{eq:confbrane} lives represents $x_{7,8,9}$. The 6d theory in the tensor branch is a quiver given by
\begin{equation}
\begin{tikzpicture}
\node (v0) at (0,-1.5) {$[Sp(2k-4)]$};
\node (v1) at (0,0) {$\stackrel{Spin(2k+4)}{4}$};
\node (v2) at (2.5,0) {$\stackrel{Sp(2k-4)}{1} $};
\node (v3) at (5,0) {$\stackrel{Spin(2k+4)}{4}$};
\node (v5) at (7.5,0) {$\stackrel{Sp(2k-4)}{1}$};

\node (v6) at (10,0) {$\stackrel{Spin(2k+4)}{4}$};
\node (v7) at (10,-1.5) {$[Sp(2k-4)]$};
\draw (v0) edge (v1);
\draw (v1) edge (v2);
\draw (v2) edge (v3);
\draw [dashed] (v3) edge (v5);
\draw (v5) edge (v6);
\draw (v6) edge (v7);
\end{tikzpicture}
\end{equation}
The 1-form symmetry is $\Gamma^{(1)}=\mathbb{Z}_2$ and it sits diagonally in all the center of the gauge groups,  and under which the vector representation of $Spin$ is not charged.
The holographic dual solutions, which are the near-horizon limit of the brane set-up \eqref{eq:confbrane}, have been discussed in \cite{Apruzzi:2013yva,Apruzzi:2017nck}, and there are two features of these solutions that are important for our purposes. The first one is that the space is AdS$_7 \times M_3$, where $M_3$ is a 3-dimensional closed manifold consisting of a $\mathbb{RP}^2$ fibered over an interval that is topologically equivalent to a 3-sphere.  The second one is that there are two D6/O6$^{+}$ sources at the two poles.
Due to the presence of the O6 plane there is a non-trivial action on the 10-dimensional NSNS B-field of IIA, $B\rightarrow -B$ \cite{Bergman:2001rp, Hanany:2000fq}. This implies that the surviving modes are the one such that 
\begin{equation}
2 B =0 
\end{equation}
so the holonomies of $B$ are integers modulo 2.  In addition the net NS5-brane charge or equivalently the $H$ flux quanta in the holographic dual are $\int_{M_3} H = 4N$.  The symmetry TFT is derived by reducing the topological action of IIA supergravity together with contributions from brane sources,  that in this case involves
\begin{equation}
S_{\rm top}^{IIA} = 2\pi \int dC_3 \wedge C_3 \wedge H + \delta({\rm sources}) S_{CS}^{O6^+/D6}
\end{equation}
where $C_p$ are the RR potentials, and the 6-brane Chern-Simons action at large $k$ D6-branes reads, 
\begin{equation}
S_{CS}^{O6^+/D6} = 2\pi \left(4k \int \sum_p C_p \wedge e^{-B} \right)
\end{equation}
where we do not consider any gauge field living on the branes, as well as any nontrivial space-time curvature and R-symmetry backgrounds, see \cite{Morales:1998ux} for the complete action. 
The part that involves the 2-form symmetry background has 2 contributions. The first one come from the bulk topological action  the second comes from the contributions of the D6 branes stacks present at the pole of the holographic solution.  At large $N$ and large $k$ we have
\begin{equation}
S_{\rm symTFT} = 2\pi \left(4N \int C_3 \wedge dC_3 + 4k \int C_3 \wedge \frac{1}{2} B \wedge B )\right).
\end{equation}
This confirms the structure of the symmetry TFT action obtained from the tensor branch, and matches at large $k$ the case \eqref{eq:NHCs} with $n=4$ and $N=1$, and $G=Spin(2k+4)$ with $k$ odd.

\section{A 5d Perspective} \label{sec:5dperp}
We will now describe how to derive the symmetry TFT in 5d and then discuss the compactification of the 6d SymTFT on a circle.  We will subsequently discuss some example of 5d theories and their SymTFTs derived in the Coulomb Branch,  which match the results obtained in \cite{Apruzzi:2021nmk}.  We shall see that in both cases,  for some specific choices of boundary conditions the SymTFT reduce to an invertible theory that reproduces the anomaly theory of the absolute 5d SCFT. 

\subsection{Derivation from 5d}
Let us now discuss how one generally couples the background field to a 5d action in the (partial) Coulomb branch.  A generic Coulomb branch action reads,
\begin{equation}
S_{CB} = 2 \pi \int \left(\frac{1}{2} \mathcal G_{ij} f^i \wedge \ast f^j + \frac{c_{ijk}}{6} a^i \wedge f^j \wedge f^ k \right) +\ldots
\end{equation}
where we do not display the scalar and fermion parts,  for $i=1,\ldots, r$.  The electric 1-form symmetry currents are given by $J_{2\, i}= \mathcal G_{ij} f^j +  \ast_5 \frac{c_{ijk}}{6} a^j f^k $.  The presence of the CS term breaks the electric $U(1)$ 1-form symmetry to a discrete subgroup depending on the levels $c_{ijk}$. This can be read off from the massive states that are decoupled and generate these CS terms via loop-corrections \cite{Bhardwaj:2020phs}. The magnetic 2-form symmetry currents are $J_3^i = \ast_5 f^i$. Similarly to Maxwell theory in 4d \cite{Cordova:2018cvg}, we can now couple the previous action to the $U(1)$ background fields for the 1-form symmetries and 2-form symmetries,  $B_2^i, B_3^i$. The action reads, 
\be
\ba
S[B_2,B_3] = 2 \pi \int  & \left( \frac{1}{2 } \mathcal{G}_{ij} (f^i - B_2^i) \wedge \ast (f^j - B_2^j)  + \mathcal G_{ij} B_3^i \wedge f^j \right.\\
 & \left. +  \frac{c_{ijk}}{6} A^i (f^j-B^j_2) (f^k-B_2^k) + \frac{1}{2}A_{\eta} \mathcal G_{ij} (f^i- B_2^i) \wedge (f^j-B_2^j) \right) 
\ea
\ee
where generically we have many emergent $U(1)$ currents due to the conservation equations $d(f^i\wedge f^j)=0$. However, in certain cases such us theories with a low-energy non-abelian gauge theory description, there is a natural coupling to $A_{\eta}$ background fields corresponding to $U(1)_{\eta}$ symmetries, where $\eta$ labels the diagonal blocks of $\mathcal{G}_{ij}$. These symmetries are the ones that contribute (sometime enhancing) to the symmetry of the UV SCFT.\footnote{In a gauge theory, $\mathcal{G}$ is the Cartan matrix of the algebra, and therefore it has a single block. $A_{\eta}$ in this case corresponds to background of the Instanton symmetry associated to the topological current $J_I=\frac{1}{4}{\rm Tr}(f\wedge f)$} In addition, the first term is invariant under the 1-form symmetry action, 
\begin{equation}
a^i \rightarrow a^i + \Lambda_{(1)}^i, \qquad B_2^i \rightarrow B_2^i +  d\Lambda_{(1)}^i
\end{equation}
whereas the second and third terms lead to a shift of the action, as well as $A_{\eta} \rightarrow A_{\eta} + \alpha_{\eta}$. All of those can be reabsorbed by, 
\begin{equation}
S_{\rm symTFT} = 2 \pi \int \left(  \mathcal G_{ij}B_3^i d B_2^j +  \frac{c_{ijk}}{6} B_2^i B_2^j B_2^k + \frac{1}{2} A_{\eta} \mathcal{G}_{ij} B_2^iB_2^j \right) 
\end{equation}
It is very important to notice that there are corrections that come from coupling the action to the first pontryagin class $p_1(TM_6)$, via the congruence discussed in \cite{Apruzzi:2021nmk},  
\begin{equation} \label{eq:congr}
    x p_1= 4 x^3 \; {\rm mod} \; 24 \qquad x \in H^2(M_6, \mathbb{Z})
\end{equation}
The modified action then reads,
\begin{equation}
S_{\rm SymTFT} = 2 \pi \int \left(\mathcal{G}_{ij} B_3^i d B_2^j +  \frac{c_{ijk}}{6} B_2^i B_2^j B_2^k + \frac{1}{24} c_i B_2^i \, p_1(TM_6)+ \frac{1}{2} A_{\eta} \mathcal{G}_{ij} B_2^iB_2^j \right) 
\end{equation} 
We will discuss later how to compute the $c_i$ coefficient in variuous models.
To manifestly see the 1-form symmetries and the anomalies,  we need to use the Smith normal form of $\mathcal G_{ij}$,  that reads,
\begin{equation}
\mathcal{G}^{\rm SNF}= P \mathcal{G} T= {\rm diag}(p_1,\ldots, p_r)
\end{equation}
where $P$ and $T$ are two $r\times r$ square matrices.  The symmetry TFT then reads,
\be
\ba \label{eq:symtftCBrot}
S_{\rm SymTFT} = 2 \pi \int& \left(p_i \tilde{B}_3^i d \tilde{B}_2^i + \frac{1}{2} A_{\eta} \mathcal{G}_{ij} ( \tilde{B}_2 P) ^i( \tilde{B}_2 P) ^j \right.\\
&\left. +   \frac{c_{ijk}}{6} (\tilde{B}_2 P)^i (\tilde{B}_2 P)^j (\tilde{B}_2P)^k + \frac{c_i}{24}   \, ( \tilde{B}_2 P) ^i \, p_1(TM_6)  \right)
\ea
\ee
where the last term is relevant by congruence \cite{Apruzzi:2021nmk}.
So for the explicit example we just need to compute the integral coefficients $p_i,  c_i,  c_{ijk}$.  

\subsection{SymTFT for 6d theories from compactification}
We now compute the compactification on $S^1$ of a given SymTFT for a general 6d SCFT disucussed in section \ref{sec:symtft6d}.  This will lead to the symmetry TFT of the 5d kk-theory that uplift to 6d in the UV.  The reduction ansatz reads
\begin{equation}
C_3^i = B_3^i + \omega \wedge \hat{B}_2^i, \qquad B_2^i = B_2^i
\end{equation}
where $\omega = d \beta - A$ with $A$ the background field for the isometry of the $S^1$ direction.  Plugging this into \eqref{eq:symTFT} and integrating over $S^1$,  we get
\begin{equation} \label{eq:symTFT2}
S_{\rm SymTFT} = 2\pi  \Omega_{ij} \int  \left( \hat B_2^i \, dB_3^j +\hat B_2^i \, \alpha_G^j \mathfrak{P}(B_2) + \frac{1}{2} dA \hat{B}^i \, \hat{B}^j  \right).
\end{equation}
For simplicity let us focus on the single tensor cases \eqref{eq:NHCs} with $\Omega_{11}=n$, 
\begin{equation} \label{eq:symTFT3}
S_{\rm SymTFT} = 2\pi  n \int  \left( \hat B_2\, dB_3 +\hat B_2 \, \alpha_G \mathfrak{P}(B_2) + \frac{1}{2} dA \hat{B}_2 \, \hat{B}_2 \right).
\end{equation}
where we recall that we choose the global structure for the gauge group such that the group is the simply connected version.  Let us study two particularly meaningful boundary conditions of this action.  The first one is given by varying the action w.r.t. $B_3$, and it reads
\begin{equation}
n d \hat{B}_2 =0 
\end{equation}
This gives rise to a absolute theory with $\Gamma^{(1)}_{\rm full}=\mathbb{Z}_n \times \Gamma^{(1)}$,  where $\Gamma^{(1)}$ is given in \eqref{eq:NHCs}.  In this case the periodicities of $\hat{B}_2$ are in $\mathbb{Z}/n$.  What is left is an invertible field TFT action that reads,
\begin{equation}
S_{\rm an} = 2\pi  n \int \left(\hat B_2 \, \alpha_G \mathfrak{P}(B_2) + \frac{1}{2} dA \hat{B}_2 \, \hat{B}_2 \right)
\end{equation}
which corresponds to the anomaly of the absolute 5d kk-theory.  The second boundary condition which is significant to study is the variation of the action w.r.t.  $\hat{B}_2$.  The boundary condition implies,
\begin{equation}
n (d B_3 + \alpha_G  \mathfrak{P}(B_2) ) =0 
\end{equation}
that is a differential operation associated to the short exact sequence $1\rightarrow \Gamma^{(2)} \rightarrow Z_3 \rightarrow \Gamma^{(1)}\rightarrow1$, and this realizes a 3-group structure for the theory at the boundary.  This structure is already present in the relative theory and implied by gauge invariance of the symmetry TFT action \eqref{eq:3group}.

\subsection{SymTFT for rank 1 6d theories from 5d} \label{sec:reduction}
For the 6d rank 1 theories compactified on $S^1$,  the constant part of $\mathcal{G}$ is given by the affine Cartan matrix of the gauge groups \eqref{eq:NHCs}, and the Chern-Simons level $c_{ijk}$ can be read off from an (affine) gauge theory prepotential,  \cite{DelZotto:2017pti,Apruzzi:2019kgb}.  We recall that the prepotential is a cubic function of the Coulomb branch scalars and it is given by,  \cite{Intriligator:1997pq,Closset:2018bjz,Hayashi:2019jvx}.
\begin{align} \label{eq:prep}
\ba
\mathcal{F}
=& \left( \frac{1}{2g_{YM}^2}\, C_{ij} \phi^i \phi^j + \frac{ \kappa}{6} \, d_{ij \ell} \phi^i \phi^j \phi^\ell \right)
 +\frac{1}{12} \left( \sum_{\alpha \in \Phi_{\mathfrak{g}}} |\alpha_i \, \phi^i|^3 - \sum_{{\bf R}_f} \sum_{ \lambda \in \mathbf W_{{\bf R}_f}} |\lambda_i \, \phi^i + m_f|^3\right),
\ea
\end{align} 
where $d_{ij\ell}= \frac{1}{2} {\rm tr}_\text{fund}\left( T_i (T_j T_\ell +
T_\ell T_j)\right)$, $C_{ij} = \text{tr} (T_i T_j)$, and $\kappa$
is the classical Chern--Simons level,  which is half-integer quantized. 
 $\Phi_{\mathfrak{g}}$ are the roots of $\mathfrak{g}$,  and $\mathbf W_{{\bf R}_f}$ are weights of the representation
${\bf R}_f$ if matter (hypermultiplets) with mass $m_f$ is present.
Then we hae 
\begin{align}
  \begin{split}
    & \mathcal{G}^{\rm full}_{ij} = \frac{\partial^2\mathcal F}{\partial \phi^i \partial \phi^j} \, ,\\
    & c_{ij \ell} = \frac{\partial^3\mathcal F}{\partial \phi^i \partial \phi^j\partial \phi^\ell} \, .
  \end{split}
\end{align}
In our case we are interested in the constant values of $ \mathcal{G}^{\rm}_{ij}=\hat{C}_{ij}$ and $c_{ij \ell}$. Finally, the $c_i$ coefficient can be computed by integrating our the fermions that are superpartner of the W-boson. The 1-loop Feynman diagram with these massive fermions runnning in the loop generates also mixed gravitational/gauge Chern-Simons terms, when the gravitational backgrounds are turned on \cite{Katz:2020ewz, Bonetti:2013cza}. 

For example if we specify $G=Spin(8)$ we get,  
\begin{equation}
\mathcal{G}=\begin{pmatrix}
2 & 0 &-1 & 0 & 0\\
0 & 2 & -1 & 0 & 0\\
-1 & -1 & 2 & -1 & -1\\
0 & 0 & -1 & 2 & 0\\
0 & 0 & -1 & 0 & 2
\end{pmatrix}
\end{equation}
The Smith normal form reads,
\begin{equation} \label{eq:so8SNFcart}
\mathcal{G}^{\rm SNF}=\begin{pmatrix}
1 & 0 &0 & 0 & 0\\
0 & 1 & 0 & 0 & 0\\
0 & 0 & 2 & 0 & 0\\
0 & 0 & 0 & 2 & 0\\
0 & 0 & 0 & 0 & 0
\end{pmatrix}, \qquad P=\begin{pmatrix}
2& 2& 3& 2& 0\\
1&1& 2& 1&0\\
1&0&2&1&0\\
0&-1&0&1&0\\
1&1&2&1&1
\end{pmatrix}
\end{equation}
We also notice that \eqref{eq:so8SNFcart} has a 0 diagonal element. This is due to the fact that in the Coulomb branch we see a $U(1) \times \mathbb{Z}_2 \times \mathbb{Z}_2$ 1-form symmetry where the $U(1)$ is broken by the massive instanton particle to $\mathbb{Z}_4$, and the $\mathbb{Z}_2$'s correspond to the left and right center symmetry of $Spin(8)$.
$c_{ijk}$ can be read off by plugging into \eqref{eq:prep} the affine root lattice.  The non-vanishing terms are,
\begin{equation} \label{eq:CSlevel}
c_{133}=c_{233}=c_{433}=c_{533}=-2, \qquad c_{111}=c_{222}=c_{333}=c_{444}=c_{555}=8. 
\end{equation}

We can now plug into \eqref{eq:symtftCBrot} the matrices \eqref{eq:so8SNFcart} as well as the CS numbers \eqref{eq:CSlevel}. The symmetry TFT then reads, \footnote{Where the upper indices are not powers but rather labels corresponding to $B^i_2$. This notation is used through the entire paper.}
\be\ba \label{eq:so8symtftfin}
    S_{\rm SymTFT}= 2\pi \int & (\tilde{B}^1_2 d \tilde{B}_3^1 + \tilde{B}^2_2 d \tilde{B}_3^2+ 2 \tilde{B}^3_2 d \tilde{B}_3^3 + 2 \tilde{B}^4_2 d \tilde{B}_3^4 + 4 \tilde{B}^5_2 d \tilde{B}_3^5 \\
      &+ 2 dA \tilde{B}^5_2 \tilde{B}^5_2 +8 \tilde{B}^5_2 \tilde{B}^3_2\tilde{B}^4_2+ 8 \tilde{B}_2^5 \tilde{B}_2^3 \tilde{B}_2^3 + 8 \tilde{B}_2^5 \tilde{B}_2^4 \tilde{B}_2^4 )
\ea\ee
where in this action we show the $\mathbb{Z}_4$ SymTFT, which is visible via a careful computation of the charge matrix as in \cite{Albertini:2020mdx, Morrison:2020ool}. This includes the charge under the center symmetries of the instanton particles, which are also visible through (mixed) Chern-Simons level in the Coulomb branch. By varying this action w.r.t. the $B_3$ fields we obtain
\begin{equation} \label{eq:BCstand}
    2 dB_2^3=0, \qquad 2 dB_2^4=0, \qquad 4 dB_2^5=0.
\end{equation}
Thus, we have that $B_2^3$ and $B_2^4$ have $\mathbb{Z}/2$ periodicities, whereas $B_2^5$ has $\mathbb{Z}/4$ periods. By choosing the boundary conditions \eqref{eq:BCstand} the second part of the action, which contains the cubic couplings, \eqref{eq:so8symtftfin} becomes an invertible theory that corresponds to the anomaly of the absolute 5d KK-theory. This also coincides with the circle reduction of the symmetry TFT for the 6d theory in the tensor branch described in the previous section \ref{sec:reduction}. Analogous results were obtained in \cite{Cvetic:2021sxm,Hubner:2022kxr} by using M/F-theory geometry.

\subsection{5d $SU(p)_q$ gauge theories}
Using the action \eqref{eq:symtftCBrot} we can also derive the symmetry TFT for 5d $SU(p)_q$ gauge theories. In order to do this we need to evaluate $\mathcal{G}$, which in this case corresponds to the Cartan of the $\mathfrak{su}_p$ gauge algebra. Whereas the $c_{ijk}$ can be computed from the prepotential \eqref{eq:prep}. The coefficient $c_i$ read \cite{Katz:2020ewz, Bonetti:2013cza},
\begin{equation}
    c_i = \frac{\partial}{\partial\phi^i}\sum_{\alpha \in \Phi_{\mathfrak{g}}} |\alpha_j \, \phi^j| 
\end{equation}
Therefore, the smith normal form $\mathcal{G}^{\rm SNF}= P \mathcal{G}T $ reads,
\begin{equation} \label{eq:supqmatrices}
    \mathcal{G}^{\rm SNF}= {\rm diag}(1,\ldots, p), \qquad P= \begin{pmatrix} 1 & 1 & \ldots &1 & 1 \\
    1 & 2 & \ldots &2 & 2 \\
    \vdots & \vdots & \ddots & \vdots & \vdots\\
     1 & 2 & \ldots &p-1 & p-1 \\
      1 & 2 & \ldots &p-1 & p \\
    \end{pmatrix} 
\end{equation}
The Chern-Simons levels have the following non-zero values,
\begin{equation} \label{eq:supqlevels}
    c_{iii}=8, \qquad c_{iii+1}= (p-2i-2) + q, \qquad c_{ii+1i+1}=(2i-p)-q, \qquad c_i=4.
\end{equation}
In the presence of a non-trivial classical Chern-Simons level the 1-form symmetry is further broken to $\Gamma^{(1)}=\mathbb{Z}_{\text{gcd}(p,q)}$ \cite{Albertini:2020mdx, Morrison:2020ool}. Plugging into \eqref{eq:symtftCBrot} the matrices, \eqref{eq:supqmatrices}, and the levels \eqref{eq:supqlevels}, we obtain
\begin{equation}
    S_{\rm SymTFT}= 2\pi \int \text{gcd}(p,q) B_2 d B_3 + \frac{qp(p-1)(p-2)}{6}B_2 B_2 B_2 + \frac{p(p-1)}{2} dA B_2 B_2 
    \end{equation}
where we use the congruence \eqref{eq:congr}. The coefficient of the first term is $ \text{gcd}(p,q)$ instead of $p$, as the classical Chern-Simons level dictates. This can be computed by including in the $\mathcal{G}$ matrix the charges of the instanton particle under the $U(1)^{(1)}$ center of the $U(1)$ Coulomb branch gauge groups, that is
\begin{equation}
    \mathcal M_{iJ} = \begin{pmatrix} 0&\\
                                 0&\\
                                 \vdots &\mathcal{G}_{ij}\\
                                 q-p-2& \\
                                 p-q& 
                                 \end{pmatrix}
\end{equation}
where $J=0,1,\dots, p$. The SNF form of this matrix provides information about the flavor symmetry rank as well. 

When varying the action w.r.t. $B_3$ we get ${\rm gcd}(p,q) dB_2=0$ and periodicites of the $B_2$ fields are then $\frac{\mathbb{Z}}{{\rm gcd}(p,q) }$.
The term $\frac{p(p-1)}{2} dA B_2 B_2$ corresponds to a mixed anomaly of the absolute theory between the $U(1)$ instanton symmetry and $\Gamma^{(1)}=\mathbb{Z}_{\text{gcd}(p,q)}$ discussed in \cite{BenettiGenolini:2020doj}. The cubic term is instead a 't Hooft anomaly for $\Gamma^{(1)}=\mathbb{Z}_{\text{gcd}(p,q)}$ studied from the non-abelian gauge theory point of view in \cite{Gukov:2020btk}. The full anomaly theory coincides with the one obtained in \cite{Apruzzi:2021nmk,BenettiGenolini:2020doj,Gukov:2020btk}.

\subsection{5d non-Lagrangian theories}
We can apply the Coulomb branch method to compute anomalies of 5d SCFTs with no non-abelian gauge theory description in the IR. These are for instance the theories analyzed in \cite{Eckhard:2020jyr, Morrison:2020ool, Apruzzi:2021nmk} called ${B_N,B_N^{(1)},B_N^{(2)}}$. Even if these theories do not have a non-abelian gauge theory description they do admit a IR Coulomb branch Lagrangian, that is fixed by enough knowledge of the electric charges, which will fix $\mathcal{G}_{ij}$, as well as the Chern-Simons level $c_{ijk}$. We can read off $\mathcal{G}_{ij}$ and $c_{ijk}$ from the intersection numbers computed via toric geometry \cite{Eckhard:2020jyr, Morrison:2020ool}. Plugging in these data \eqref{eq:symtftCBrot} we are able to reproduce the expression in \cite{Apruzzi:2021nmk}. Let us discuss how this work in an explicit example, such us $B_3$, where from toric geometry we can read
\begin{equation}
    \mathcal{G}_{11}=3, \qquad c_{111}=9, \qquad c_1=1.
\end{equation}
The coefficient $c_1$ geometrically corresponds to the intersection between the compact surface class $S$ and the second Chern-class of the toric Calabi-Yau \cite{Bonetti:2013cza}. This implies the following symmetry TFT action,
\begin{equation}
    S_{\rm SymTFT}= 2\pi \int 3 B_2 d B_3 + \frac{3}{2}B_2 B_2 B_2 + \frac{1}{24} B_2 p_1(TM_6)  
    \end{equation}
    By varying with respect of $B_3$ and in terms of integral periodicity fields $\frac{\hat{B}_2}{3}=B_2$, we get an anomaly action that is
    \begin{equation}
         S_{\rm Anomaly}=2\pi \int  \frac{1}{9} \hat{B}_2 \hat{B}_2 \hat{B}_2
    \end{equation}
    where we used the congurence \eqref{eq:congr}.

\section{Fate of 2-form symmetries in 6d supergravity}\label{sec:sugra}
Let us finally comment on what happens to higher-form symmetries of the 6d rank-1 theories when we consistently couple them to 6d dynamical gravity. As anticipated this can be done consistently because there exist string theory constructions realizing this coupling. The theories obtained via this construction are called non-Higgsable clusters (NHCs), which are engineered in F-theory via compactification on Calabi-Yau elliptically fibered over Hirzebrouch surfaces $\mathbb{F}_n$ \cite{Morrison:2012np}. The intersection pairing reads,
\begin{equation}   \label{eq:sugrapair}
    \Omega= \begin{pmatrix}  0 & -1\\
     -1 & n \end{pmatrix}
\end{equation}
with $n>2$. This consists of coupling the theories in \eqref{eq:NHCs} to a self-dual tensor field which is present in the gravity multiplet, where the intersection pairing is given above. The breaking of the 2-form symmetry $\Gamma^{(2)}=\mathbb{Z}_n$ associated to the antiself-dual tensor with self-pairing $n$ is given by a mechanism similar to the one described in \cite{Apruzzi:2020zot}. For instance, we can check whether backgrounds for $\Gamma^{(2)}$, $C_3^2$, such that the periodicities of $dC_3$ are fractional and they have $\mathbb{Z}/n$ periods, are consistent. In the effective action we have the coupling $\Omega_{ij} b^i \, dC_3^j = b_1 \, dC_3^2 $, and the Bianchi identity reads. This coupling measures the induced charged by the background fields on the BPS strings, and therefore, because the BPS string lattice needs to be integrally quantized by Dirac quantization, it must have integer coefficient if the integrand has integral periods. 
Thus, we conclude that the activation of the background field for $\Gamma^{(2)}=\mathbb{Z}_n$ violates Dirac quantization of the BPS string charges, charged under the self-dual tensor of the gravity multiplet in 6d, also called supergravity strings. Our proposal provides a field theory explanation of the string theoretic/geometric observation in \cite{Braun:2021sex}. Another equivalent way of phrasing this mechanism is that the 2-form symmetry is broken since the surface defect are screened by the supergravity strings. This can also be deduced by computing the Smith normal form of \eqref{eq:sugrapair}, which result in a trivial 2-form symmetry. Since the supergravity strings are tensionless only in the UV (or at infinite distances), and do not lead to any massless state at low-energy the symmetries can be treated as approximate in the sense of \cite{Cordova:2022rer}, whereas they are broken at a energy below the Plank scale. 

Another interesting model discussed in \cite{Apruzzi:2020zot} is given is specified by the following intersection pairing,
\begin{equation}   
    \Omega= \begin{pmatrix}  0 & -1&-5\\
     -1 & 3&0 \\
     -5 & 0 & -75\end{pmatrix}
\end{equation}
where we have an $SU(3)$ associated with the tensor multiplet of self pairing $3$ and another $SU(3)$ on the one with self pairing $-15$ with no massless matter. In the tensor branch action coupled to backgrounds we have the following relevant coupling terms
\begin{equation}
    S \supset 2 \pi \int - b_1 \,  (\alpha_{SU(3)} P(B_2^1) + 5 \alpha_{SU(3)}P(B_2^3)+ dC_3^1 + 5 dC_3^3) + \ldots)
\end{equation}
where $b^i$ are dynamical 2-form tensor fields.
This couplings tell us that the diagonal 1-form symmetry $\mathbb{Z}_3^{\rm diag}\in Z(SU(3)\times SU(3))$, whose background is $B_2^1=B_2^3$, together with a $\Gamma^{(2)}=\mathbb{Z}_3$, with background $C_3^1=C_3^3$, symmetry background do not violate Dirac quantization for the supergravity strings.
This means that the entire 3-group symmetry should be gauged combining the geometric evidence of the constructions in both \cite{Apruzzi:2020zot} and \cite{Braun:2021sex}.

\section*{Acknowledgements}
We thank Pietro Benetti Genolini, Ling lin, Sakura Sch\"afer-Nameki, Yinan Wang for discussions and correspondence. 
FA is supported by the Albert Einstein Center for fundamental physics at Bern university, the Swiss
National Science Foundation and in part by the European Union’s Horizon 2020 Framework: ERC grant 682608.

\appendix

\bibliographystyle{JHEP}
\bibliography{F}

\end{document}